\documentclass[layout=twocolumn, manuscript=letter, journal=nalefd]{achemso}

\usepackage[utf8]{inputenc}
\usepackage[T1]{fontenc}
\usepackage{amsmath}
\usepackage{graphicx}
\usepackage[english]{babel}
\usepackage{float}
\usepackage[font=small,labelfont=bf]{caption}
\usepackage{color}
\usepackage{multicol}
\usepackage[version=3]{mhchem} 

\DeclareUnicodeCharacter{0303}{\~{}}
\DeclareUnicodeCharacter{0300}{\`{}}

\author{José Roberto de Toledo}
\email{jose.toledo@df.ufscar.br}

\author{Caique Serati de Brito}
\affiliation{Physics Department, Federal University of S\~ao Carlos, 13565-905 S\~ao Carlos, SP  Brazil}

\author{Barbara L. T. Rosa}
\affiliation{Institut für Festkörperphysik, Technische Universität, 10623 Berlin, Germany}

\author{Alisson R. Cadore}
\affiliation{Brazilian Nanotechnology National Laboratory (LNNano), Brazilian Center for Research in Energy and Materials (CNPEM), 13083-100 Campinas, SP, Brazil}
\alsoaffiliation{Programa de Pós-Graduação em Física, Instituto de Física, Universidade Federal do Mato Grosso, 79070-900 Cuiabá , Brazil}

\author{César Ricardo Rabahi}
\affiliation{Physics Department, Federal University of S\~ao Carlos, 13565-905 S\~ao Carlos, SP  Brazil}

\author{Paulo E. Faria Junior}
\affiliation{Department of Physics, University of Central Florida, Orlando, 32816 Florida, USA}
\alsoaffiliation{Department of Electrical and Computer Engineering, University of Central Florida, Orlando, 32816 Florida, USA}

\author{Ana Carolina Ferreira de Brito}
\affiliation{Brazilian Synchrotron Light Laboratory (LNLS), Brazilian Center for Research in Energy and Materials (CNPEM), 13083-100 Campinas, SP, Brazil}

\author{Talieh S. Ghiasi}
\author{Josep Ingla-Aynés}
\affiliation{Kavli Institute of Nanoscience, Delft University of Technology, 2628 CJ Delft, The Netherlands}

\author{Christian Sch\"uller}
\affiliation{Institut für Experimentelle und Angewandte Physik, Universit\"at Regensburg, D-93040 Regensburg, Germany}

\author{Herre S. J. van der Zant}
\affiliation{Kavli Institute of Nanoscience, Delft University of Technology, 2628 CJ Delft, The Netherlands}

\author{Stephan Reitzenstein}
\affiliation{Institut für Festkörperphysik, Technische Universität, 10623 Berlin, Germany}

\author{Ingrid D. Barcelos}
\affiliation{Brazilian Synchrotron Light Laboratory (LNLS), Brazilian Center for Research in Energy and Materials (CNPEM), 13083-100 Campinas, SP, Brazil}

 \author{Florian Dirnberger}
\affiliation{Department of Physics, TUM School of Natural Sciences, Technical University of Munich, 85748 Garching, Germany}
\alsoaffiliation{Zentrum für Quantum Engineering (ZQE), Technical University of Munich, 85748 Garching, Germany}
\alsoaffiliation{Munich Center for Quantum Science and Technology (MCQST), Technical University of Munich, 85748 Garching, Germany}

\author{Yara Galvão Gobato
}
\affiliation{Physics Department, Federal University of S\~ao Carlos, 13565-905 S\~ao Carlos, SP  Brazil}
\email{yara@ufscar.br}

\title{Interplay of energy and charge transfer in WSe$_2$/CrSBr heterostructures}


\begin{document}
\pagenumbering{arabic}
\newcommand{\wse}{\mathrm{WSe}_{2}}
\newcommand{\sample}{\mathrm{WSe}_{2}/\mathrm{CrSBr}}

\begin{abstract}
 Van der Waals heterostructures (vdWHs) composed of transition-metal dichalcogenides (TMDs) and layered magnetic semiconductors offer great opportunities to manipulate exciton and valley properties of TMDs. Here, we present magneto-photoluminescence (PL) studies in a $\wse$ monolayer (ML) on a CrSBr crystal, an anisotropic layered antiferromagnetic semiconductor. Our results reveal unique behavior of each of the ML-$\wse$ PL peaks under magnetic field that is distinct from the pristine case. An intriguing feature is the clear enhancement of the PL intensity that we observe each time the external magnetic field tunes the energy of an exciton in CrSBr into resonance with one of the optical states of $\wse$. This result suggests a magnetic field-controlled resonant energy transfer (RET) beyond other effects reported in similar structures. Our work provides deep insight on the importance of different mechanisms into magnetic vdWHs and underscores its great potential for light harvesting and emission enhancement of two-dimensional materials.

\end{abstract}


\section{}
Two-dimensional (2D) magnetic materials have attracted great attention in the last years, offering a new platform to study fundamental properties of magnetism in low dimensions and for possible applications in spintronics\cite{gong2017discovery,gibertini2019magnetic,jiang2021recent,wang2022magnetic,huang2017layer,ziebel2024crsbr,long2023intrinsic,glazov2024excitons}. Among those, CrSBr has received increasing attention in the last few years because of its inherently coupled magnetic and optical properties \cite{ziebel2024crsbr,long2023intrinsic,wilson2021interlayer,glazov2024excitons}. CrSBr is an air-stable quasi-1D  van der Waals (vdW) semiconductor material with direct band gap of about 1.5 eV \cite{ziebel2024crsbr,smolenski2024large,nessi_polaritons_2024}. It has an orthorhombic crystal structure with a rectangular unit cell in the $\hat{a}-\hat{b}$ plane stacked along the $\hat{c}$ direction\cite{ziebel2024crsbr}. The CrSBr monolayer (ML) is ferromagnetic (FM) \cite{ziebel2024crsbr}. The interlayer exchange coupling between the layers, favors an A-type antiferromagnetic order (AFM) with a Néel temperature of approximately 135 K \cite{klein2023bulk,klein2022sensing,telford2022coupling,lopez2022dynamic,ye2022layer,bae2022exciton,long2023intrinsic,ghiasi2021electrical,dirnberger2023magneto,pawbake2023magneto}. Its electronic band structure and consequently the energy of excitons are both very sensitive to the interlayer magnetic exchange interaction which allows probing its magnetic order by magneto-optical spectroscopy\cite{wilson2021interlayer,serati2023charge}. CrSBr has two anisotropic emissions related to the fundamental bright exciton (labeled A exciton) and to a higher energy exciton (labeled B exciton) at around 1.36 eV and 1.76 eV, respectively \cite{long2023intrinsic,wilson2021interlayer,serati2023charge,pawbake2023magneto,datta2024magnon,nessi_polaritons_2024}. The A exciton is tightly bound while the B exciton has a decreased spatial localization\cite{nessi_polaritons_2024}. Furthermore, both excitons show a redshift in their PL peak energy with increasing magnetic field up to a field-induced FM state \cite{long2023intrinsic,wilson2021interlayer,serati2023charge,datta2024magnon,komar2024colossal,shi2024giant}. 

Very interesting material design opportunities appear when 2D magnetic materials can also be combined with non-magnetic materials such as ML-TMD to modify their exciton and valley properties using magnetic exchange interaction and charge transfer effects \cite{scharf2017magnetic, sierra2021van,seyler2018valley,ciorciaro2020observation,choi2023asymmetric,norden2019giant,glazov2024excitons}. Several previous studies were performed in different vdWH composed of 2D FM materials with perpendicular magnetization such as CrBr$_3$ and CrI$_3$ and ML-TMDs and have revealed important changes in the properties of ML-TMDs such as valley splitting and polarization degree under zero magnetic field \cite{seyler2018valley,ciorciaro2020observation,choi2023asymmetric,norden2019giant,lyons2020interplay}. Recent studies were focused on a ML-MoSe$_2$/CrSBr, a type-III vdWH, and have revealed important modifications in their physical properties, which were associated with magnetic proximity and charge transfer effects \cite{serati2023charge,beer2024proximity}.

There is also an increasing interest in 2D materials such as ML-WSe$_2$ to generate single-photon emitters (SPEs) for possible applications in photonic quantum technologies \cite{palacios2017large, linhart2019localized,azzam2021prospects,serati2024probing, alapatt2025highly}. Particularly, several 2D magnetic materials (Cr$_2$Ge$_2$Te$_6$, CrI$_3$ and NiPS$_3$) have been used  as substrates to modify the optical properties of SPEs evidencing an enhancement of the g-factors and circular polarization, offering an exciting platform to design quantum devices for implementing photonic quantum networks\cite{shayan2019magnetic,li2023proximity,mukherjee2020observation}. In this context, the investigation of the ML-$\wse$/CrSBr could reveal new opportunities to modify the optical properties of localized excitons in WSe$_2$ in advanced quantum light sources.
 
Here, we investigate excitonic properties of ML-$\wse$ on CrSBr using low temperature magneto-PL techniques. Different contributions of the CrSBr layer are observed in the magnetic field dependence of PL peaks in the $\wse$/CrSBr heterostructure.  In addition to charge transfer effects, we observe clear PL signatures each time the B exciton (labeled X$_B$) of CrSBr comes into resonance with an exciton  state in the ML-WSe$_2$. As the PL peak energies of CrSBr can be controlled by the applied magnetic field, the resonant condition of the energy states of CrSBr and ML-$\wse$ can be tuned by this external parameter resulting in changes in the PL properties of ML-$\wse$/CrSBr. Remarkably, the PL of the sharp emission peaks in ML-$\wse$/CrSBr shows significant intensity enhancement after the field-induced ferromagnetic state of CrSBr. Moreover, the PL intensity of excitons and trions in ML-WSe$_2$ is also enhanced at different out-of-plane magnetic fields. We suggest that these results could be explained by resonant energy transfer (RET) effect involving X$_B$ in the CrSBr. Our studies point out  the importance of different mechanisms to manipulate optical properties of excitonic states of ML-TMDs in magnetic vdWHs. 


\begin{figure*}[t]
\centering
\includegraphics[width=1\textwidth]{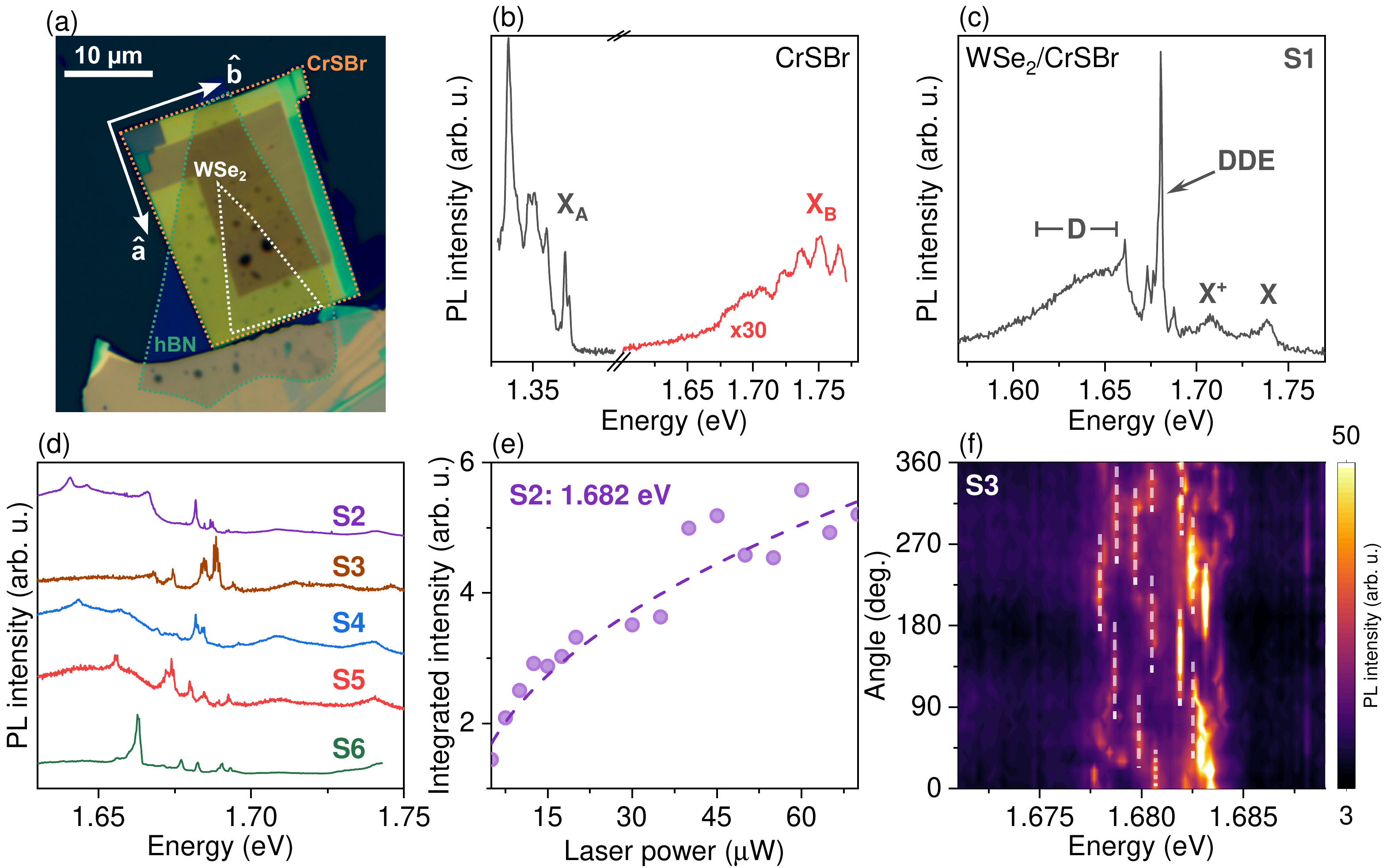}
\caption{(a) Optical microscopy image of the ML-$\wse$/CrSBr sample, indicating the orientation of the CrSBr crystallographic axes $\hat{a}$ and $\hat{b}$. (b) Typical PL spectrum of bulk CrSBr, showing the emission of the A- (black curve) and B-excitons (red curve). (c) PL spectrum of the $\sample$ heterostructure, showing several emission peaks from the WSe$_2$ layer. (d) PL spectra for different laser positions, labeled S2-S6, showing several sharp PL peaks (labeled DDE). (e) Laser power dependence of the PL intensity for a sharp emission peak at 1.682 eV. (f) Color-coded map of the PL intensity of the DDE states as a function of the in-plane polarization angle, revealing several doublet peaks. All PL data were obtained using linearly polarized laser excitation along the $\hat{b}$-axis with energy of 1.88 eV at 3.6 K.
}
\label{pl}
\end{figure*}

Our sample consists of a ML-WSe$_2$ on bulk CrSBr, capped by a thin layer of hexagonal boron nitride (hBN).  Figure \ref{pl}(a) presents an optical microscope image of the sample showing the crystal orientations, $\hat{a}$ and $\hat{b}$, of the CrSBr crystal. Figures \ref{pl}(b), (c) and (d) present the typical PL spectra for the pristine CrSBr and WSe$_2$/CrSBr for a laser excitation energy of 1.88 eV at 3.6 K. Several emission peaks are observed below 1.4 eV and associated with the A exciton (labeled X$_A$) in CrSBr \cite{ziebel2024crsbr, klein2023bulk, serati2023charge}. Additionally, we also detect a much weaker emission band in the range of 1.60 to 1.77 eV (Figure \ref{pl} (b)) which is attributed to the X$_B$ exciton in CrSBr.

Figures \ref{pl}(c) and \ref{pl}(d) show the emission peaks from the WSe$_{2}$/CrSBr which are close to the spectral range of the emission of the $X_B$ in CrSBr. The bright exciton (X) and trion (X$^{+}$) PL peaks of ML-WSe$_2$ are observed at around 1.738 eV and 1.707 eV, respectively. Furthermore, we also observe a broad PL band that is associated with the emission of defects and labeled as defect band (D) in panel (c). Several sharp emission peaks,  are also revealed below 1.68 eV for different sample positions as shown in Figure \ref{pl}(d). As expected, the PL spectra change depending on laser position due to the presence of different local strain\cite{palacios2017large,linhart2019localized,chen2019entanglement,azzam2021prospects,de2022strain,cavalini2024revealing,glazov2024excitons,de2022strain,serati2024probing}. They are associated with defect dark exciton states (DDE) \cite{linhart2019localized} which are promising candidates of single-photon emitters (SPEs). 
 Figure \ref{pl}(e) shows the integrated PL intensity of one of the sharp PL peaks as a function of the laser power. A saturation trend is observed which clearly confirms the localized nature of these emission peaks \cite{kern2016nanoscale, branny2017deterministic, palacios2017large, parto2021defect, blundo2024alice, linhart2019localized,serati2024probing}. Figure \ref{pl}(f) shows a typical color-coded map of the linearly-polarized emission intensity of the DDE peaks as a function of the angle of in-plane polarization. We evidence several doublet emission peaks with orthogonal linear polarization, showing a  typical zero-field splitting of $\delta \approx 0.65$ meV (for details see equation (S1) in SI file) similar to previous reports in the literature\cite{robert2017fine, srivastava2015optically, he2015single, chakraborty2015voltage, koperski2017optical, ren2019review, michaelis2022single, kumar2015strain, palacios2017large,chen2019entanglement}.

\begin{figure*}
\centering
\includegraphics[width=1\textwidth]{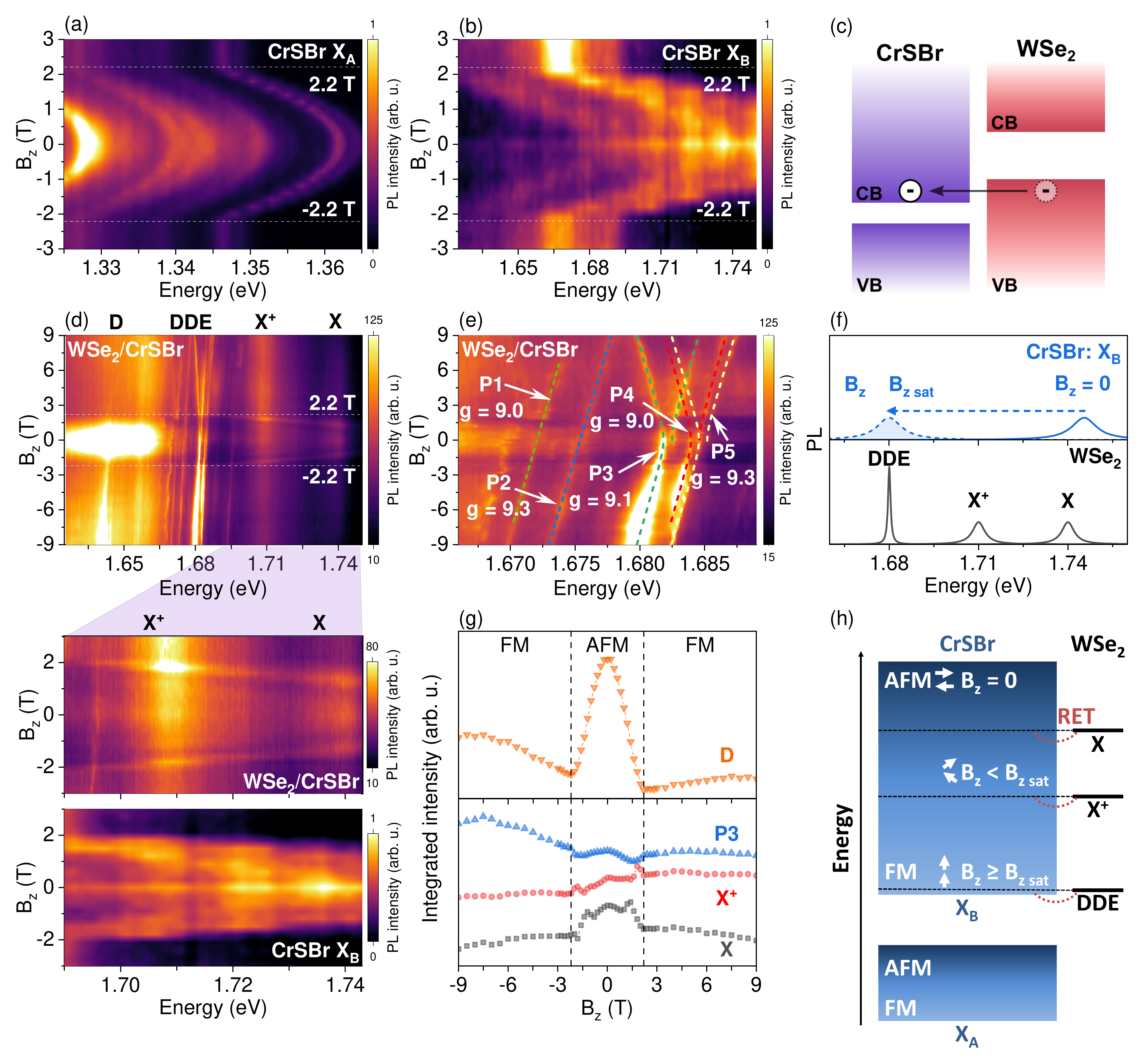}
\caption{(a,b) Color-coded map of the circularly polarized PL intensity for the fundamental (X$_A$) and X$_B$ in pristine CrSBr flake respectively, as a function of a magnetic field applied along the $\hat{c}$-axis (labeled B$_z$). (c) Schematic representation of the type-III band alignment for the WSe$_2$/CrSBr heterostructure showing the charge transfer effect between the layers. (d) Color-coded map of the circularly polarized PL spectra in WSe$_2$/CrSBr heterostructure as a function of magnetic field. Magnified view of the color-coded map showing anomalous changes in the PL intensity for ML WSe$_2$ due to resonant condition of excitonic states in ML WSe$_2$ and X$_B$ in CrSBr. (e) Enlarged view of the color-coded map in the spectral range corresponding to the emission of the localized excitons in panel (d). The dashed lines indicate the magnetic field dependence of the PL peak energies as a function of the magnetic field. (f) Schematic representation of the pristine CrSBr and ML-WSe$_2$ PL bands, showing that by applying a magnetic field, the emission of the X$_B$ can be tuned into resonance with the emissions of WSe$_2$ in the heterostructure. (g) Magnetic field dependence of the integrated PL intensity for the D band, X, X$^{+}$ and one of the DDE emission peak (P3 emission on panel (e)). The vertical dashed lines indicate the saturation magnetic fields of CrSBr ($|B_{z\ sat}|\approx 2.2$ T). (h) Schematic drawing of the  a possible RET effect tuned by magnetic field in the WSe$_2$/CrSBr heterostructure. The PL data were obtained using a linearly polarized laser along $\hat{b}$-axis with energy of 1.88 eV. The $\sigma^-$ PL component was collected for positive magnetic field at 3.6 K}
\label{faraday_color}
\end{figure*}

In order to investigate the impact of the CrSBr layer on the exciton and valley properties of ML-WSe$_2$, we have performed circularly polarized resolved PL measurements under out-of-plane magnetic field (B$_z$). Figures \ref{faraday_color}(a) and \ref{faraday_color}(b) show the color-coded map of the circular polarization resolved PL intensity of the pristine CrSBr as a function of B$_z$, using a linearly polarized laser excitation parallel to the $\hat{b}$ axis of CrSBr at 3.6 K. The CrSBr PL peak energy of the X$_A$ shows a redshift of about 16 meV (Figure S9 in SI) after the magnetic field induced phase transition of CrSBr with a saturation field of $|B_{z\ sat}|\approx2.2$ T. On the other hand, the X$_B$ emission (Figure \ref{faraday_color}(b)) features a higher energy redshift of about 80 meV (Figure S9) with increasing magnetic field, which is similar to very recent reports in the literature \cite{datta2024magnon,komar2024colossal,shi2024giant}. Figure 2(c) presents the schematic view of the type III-band alignment of the heterostructure to illustrate the charge transfer effect which can affect the PL intensity of WSe$_2$/CrSBr.

Figure \ref{faraday_color}(d) shows the color-coded map of the circular polarization resolved PL intensity of the WSe$_{2}$/CrSBr heterostructure as a function of B$_z$. We observe different modifications in the PL intensity for the emission peaks in ML-WSe$_2$/CrSBr  as compared to previous magneto-PL studies in MoSe$_{2}$/CrSBr \cite{serati2023charge}. In this previous study, a reduction of PL intensity of the MoSe$_{2}$ trion/exciton peak accompanied by a change of relative intensity of trion/exciton was observed after the magnetic phase transition of CrSBr\cite{serati2023charge}. On the other hand, for the  WSe$_{2}$/CrSBr heterostructure, a clear enhancement of the PL intensity of the  WSe$_{2}$ X and X$^{+}$ PL peaks is observed for magnetic fields that tune X$_B$ in CrSBr into resonance with these peaks in WSe$_{2}$ (for B$_z$ < B$_{z\ sat}$), suggesting a possible contribution of the RET effect. These effects are shown in more detail in the enlarged views of the color-coded map for the PL intensity, evidencing a clear correlation between changes of PL intensity  of WSe$_2$ with the X$_B$ PL peak energy. For  B$_z$ > B$_{z\ sat}$ the changes in the PL intensity  are dominated by the valley Zeeman effect and thermalization of carriers in the K and K´ valleys{\cite{glazov2024excitons,koperski2015single,li2014valley,de2024ultrathin,chen2018coulomb}. Furthermore, a significant enhancement of the DDE PL peaks is clearly revealed at B$_{z\ sat}$. On the contrary, the PL intensity of the D band, that is in the spectral range that cannot be tuned with X$_B$, is reduced after the phase transition in CrSBr and this effect is also explained by changes in the degree of charge transfer\cite{serati2023charge}. Figure 2(g) shows the magnetic field dependence of integrated PL intensity for D, P3, X and  X$^{+}$ emission peaks. All these results indicate a possible contribution of an additional mechanism for PL intensity of WSe$_{2}$,  for magnetic fields that tune X$_B$ \cite{shi2024giant,komar2024colossal, datta2024magnon} into resonance condition with excitonic states in WSe$_{2}$, such as a RET effect. The schematic in Figure \ref{faraday_color}(f) provides an overview of the PL of pristine CrSBr and ML-WSe$_2$ for B$_z$ = 0 T and B$_z$ > B$_{z\ sat}$. The PL energy of the X$_B$ in CrSBr is near the PL energy of pristine ML-WSe$_2$. Applying an external magnetic field, allows us to tune the X$_B$ of CrSBr into resonance with different exciton states in WSe$_2$  which could  modify the PL intensity in the ML-WSe$_2$/CrSBr heterostructure.

In order to explore the nature of the sharp PL peaks, we have also analyzed the magnetic field dependence of the energy peaks. Figure \ref{faraday_color}(e) presents a zoom of the color-coded map of the magnetic field dependence of the PL intensity for the WSe$_2$/CrSBr heterostructure in the range of sharp PL peaks (labeled P1, P2, P3, P4, and P5) and their extracted g-factor values (for details see Figure S11 and equation (S1) in SI). We observe clear evidence of doublet structures for several sharp PL peaks in agreement with our interpretation of DDE \cite{robert2017fine, de2022strain,serati2024probing}. Additionally, the extracted g-factor values of these sharp peaks are $|g|\approx9$, also in agreement with this interpretation \cite{linhart2019localized,serati2024probing}. Moreover, we remark that the circular polarization degree (CPD) of the bright X and X$^{+}$  and sharp emission peaks have opposite signs (Figure \ref{faraday_color}(d)) which suggests that they could be related to localized positively charged dark trions\cite{li2019direct, prando2021revealing,covre2022revealing,zhang2017magnetic}.In addition, the CPD for all emission peaks (Figure S10) shows important changes around the RET condition.

\begin{figure*}
\centering
\includegraphics[width=0.95\textwidth]{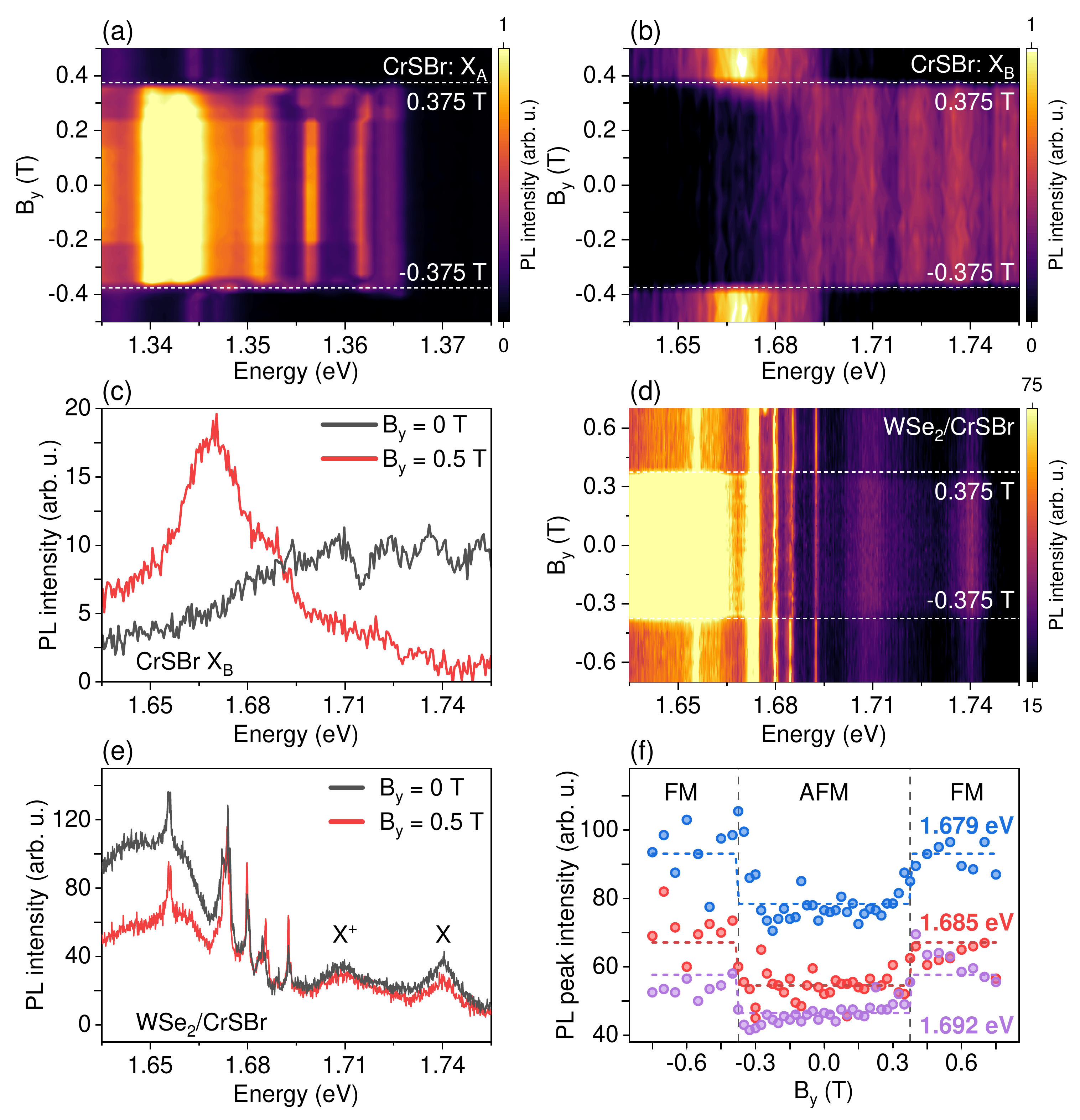}
\caption{(a,b) Color-coded map of the circularly polarized PL intensity of the A and B excitons respectively in the pristine CrSBr layer as a function of  magnetic field applied along the $\hat{b}$-axis (labeled B$_y$). (c) Typical PL spectra of B exciton of CrSBr before and after its magnetic phase transition. (d) Color-coded map of the circularly polarized PL intensity of WSe$_{2}$/CrSBr as a function of parallel magnetic field.(e) PL spectra of WSe$_2$/CrSBr before and after the magnetic phase transition of CrSBr. (f) Integrated PL intensity of selected DDE PL peaks in WSe$_{2}$/CrSBr as a function of magnetic field. The PL data were obtained using a laser with energy 1.88 eV linearly polarized along the $\hat{b}$-axis and at 3.6 K.}
\label{voigt}
\end{figure*}

To obtain a deeper understanding of the properties of our heterostructure, we have additionally measured magneto-PL with the laser excitation polarized along the $\hat{a}$-axis (Figure S14). At this condition, the PL intensity of CrSBr is much weaker due to the anisotropic properties of CrSBr. Similar results were also observed, which are also consistent with a possible RET effect. In general, these experimental results indicate a possible contribution of a magnetic field controlled interlayer RET effect involving the X$_B$ in CrSBr and excitonic states in WSe$_2$ in the heterostructure. Figure \ref{faraday_color}(h) shows a schematic drawing of the RET tuned by the magnetic field in the ML-WSe$_{2}$/CrSBr heterostructure. Depending on the magnetic field,  the RET effect modifies a different PL peak in ML-WSe$_2$. 

There are different types of RET, such as F\"{o}rster \cite{kozawa2016evidence,selig2019theory} and Dexter coupling \cite{zheng2022dexter}. Both of them were observed in several systems including type-II TMD vdWHs \cite{kozawa2016evidence,selig2019theory,zheng2022dexter}. However, there is no previous report about energy transfer in vdWHs composed of TMD and magnetic 2D materials. Particularly, the F\"{o}rster type coupling occurs only with bright emissions which have in-plane dipole momentum \cite{kozawa2016evidence,selig2019theory} and it is limited to a length scale < 10 nm. As dark excitons have out-of-plane dipole momentum \cite{luo2020exciton}, they do not allow an efficient F\"{o}rster type RET effect. However, weakly localized defective dark excitons in ML-WSe$_2$, usually observed at around 1.68 eV, have shown an in-plane dipole moment which could favor the RET effect \cite{luo2020exciton} in the WSe$_2$/CrSBr heterostructure. On the other hand, for a Dexter type of RET effect, there is no requirement to have bright emissions as it is not related to the dipole interactions. Actually, Dexter type RET depends on the overlap of their electron wavefunctions, and therefore requires that the two layers must be closely contacted (< 1 nm) \cite{luo2020exciton, selig2019theory, hu2020manipulation,zheng2022dexter}. 
Particularly, F\"{o}rster type RET would be more likely, since it is a dipole-dipole coupling, and is independent of the spin direction.  Dexter type RET relies on the wave-function overlap which is expected to be more suppressed because of the perpendicular spin directions in both materials. However, as it was previously reported that the magnetization direction of CrSBr has a perpendicular component \cite{beer2024proximity} due to magnetic proximity effects in a TMD/CrSBr heterostructure, it could allow a possible contribution of Dexter type RET for the WSe$_2$/CrSBr heterostructure. 

In order to understand our results in more detail, we have also performed time-resolved PL (TRPL) measurements (see Figure S23 in the SI file). We observed a reduction of PL time decay in the resonant energy condition. For a Förster type RET,  the acceptor should have a longer decay times. However, the assignment of which material is the acceptor and which is the donor in our heterostructure is not straightforward. In addition, we also prepared a sample CrSBr/hBN/WSe$_2$/hBN (see Figure S21) with a thin layer of hBN  (of about 2 nm) between WSe$_2$ and CrSBr. We observed that both effects (charge transfer and RET) were suppressed (see Figure S22) which indicates that they occur at a short distance (< 2 nm). This result could suggest that the RET effect is strongly dependent on the overlap of the wave functions of carriers in the WSe$_2$ and CrSBr layers, as expected for Dexter type RET. However,the detailed mechanism for the RET in the WSe$_2$/CrSBr is still unknown. Therefore, additional studies are necessary to fully understand the nature of the RET in the WSe$_{2}$/CrSBr heterostructure. 

We have also performed circular polarization resolved PL measurements under parallel magnetic fields $\vec{B}\parallel\hat{b}$ (B$_y$).  Figures \ref{voigt}(a) and \ref{voigt}(b) show the color-coded map of the PL intensity as a function of B$_y$ for X$_A$ and X$_B$ in pristine CrSBr flake showing the magnetic field induced phase transition of CrSBr at around $|B_{y\ sat}| \approx 0.375$ T. Figure \ref{voigt}(c) presents the typical PL spectra for the X$_B$ in CrSBr showing the higher redshift of the PL band after the CrSBr magnetic field phase transition. 

Figure \ref{voigt} (d) presents the color-coded map of the PL intensity for the ML-WSe$_2$/CrSBr as a function of magnetic field. Remarkably, the PL intensity of some sharp emission peaks shows an abrupt enhancement while the other PL peaks, that are not resonant with the X$_B$, show an abrupt reduction of PL intensity for B$_y$ > B$_{y\ sat}$. This PL intensity enhancement is illustrated in the selected PL sharp peaks before and after the magnetic phase transition (Figure \ref{voigt}(e)) and also in the magnetic field dependence of the PL intensity of these sharp PL peaks (Figure \ref{voigt}(f)) in WSe$_{2}$/CrSB. In contrast to the condition of out-of-plane magnetic field, the exciton/trion peaks also show a decrease in PL intensity. This occurs because under parallel magnetic field there is no resonance condition for trion/exciton and X$_B$ and the PL intensity is dominated by charge transfer effects. This PL intensity enhancement is associated with a possible contribution of RET effect while the PL decrease is associated with changes in the degree of charge transfer.

\begin{figure*}[t]
\centering
\includegraphics[width=1\textwidth]{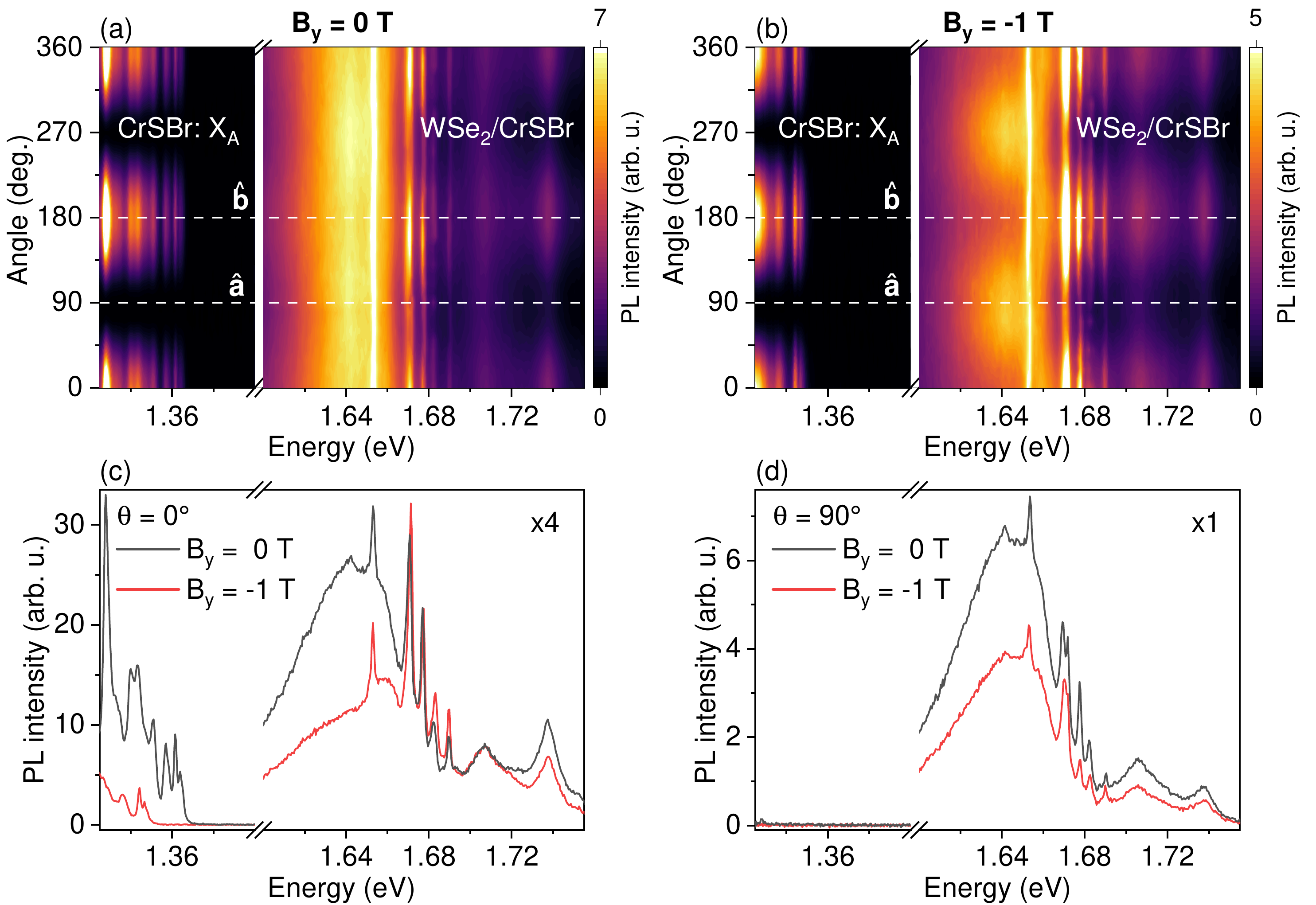}
\caption{(a,b) Color-coded map of the linearly polarized emission intensity as a function of the angle of in-plane linear polarization for the $\sample$ heterostructure at 0 T and after the magnetic phase transition (-1 T) respectively at 3.6~K. The magnetic field was applied along the $\hat{b}$-axis of CrSBr (B$_y$). (c,d) PL spectra at 0 T and -1 T for polarization angles of 0° and 90°, respectively, which suggests a possible anisotropic RET effect. The laser is linearly polarized along the $\hat{b}$ axis.}
\label{linear_pol}
\end{figure*}

 To shed more light on the impact of the anisotropic properties of the CrSBr material on the RET, we also performed linear polarization resolved PL measurements under a parallel magnetic field before and after the field-induced ferromagnetic state of the CrSBr. The use of anisotropic materials are expected to provide additional degrees of freedom for the directional control of the RET effect in vdWHs \cite{nayem2023anisotropic}. Figure 4(a) and (b) present the color-coded maps of the linearly polarized PL intensity as a function of the angle of in-plane linear polarization under 0 T and -1 T at 3.6~K ($\vec{B}\parallel\hat{b}$). Figure 4(c) and (d) show the PL spectra under 0~T and -1~T for linear polarization detection along the $\hat{a}$-axis (90°) and $\hat{b}$-axis (0°)  of CrSBr. We clearly observe that the CrSBr PL peaks are strongly linearly polarized along the $\hat{b}$ axis as expected due to the anisotropic properties of CrSBr \cite{wilson2021interlayer,serati2023charge,ziebel2024crsbr}. Interestingly, we also observed that the PL intensity of the emission peaks in WSe$_2$/CrSBr depends on the in-plane linear polarization angle. Moreover, we also observed a clear enhancement of PL intensity of the sharp emission peaks after the magnetic phase transition of CrSBr ($|B_y|>0.375$ T) and only for linearly polarization detection along $\hat{b}$ axis probably due to the anisotropic properties of CrSBr. This result is also consistent with our interpretation of the RET effect to explain the brightening of the sharp emissions. In general, our ﬁndings suggest an anisotropic RET effect which could dominate over the charge transfer effect for this WSe$_2$/CrSBr heterostructure.

In summary, we have investigated excitonic properties in a WSe$_2$/CrSBr heterostructure using linearly and circularly polarized photoluminescence under parallel and out-of-plane magnetic fields.  Our results demonstrate important modifications in the excitonic properties of ML-WSe$_{2}$ under increasing magnetic field induced by the adjacent CrSBr layer. A clear PL intensity enhancement is observed for emission peaks in WSe$_2$, under external magnetic fields that tunes the X$_B$ in CrSBr into resonance with emission peaks in WSe$_2$. This effect is associated with a possible contribution of an RET effect, controlled by the magnetic field and involving excitonic states in CrSBr and  ML-WSe$_2$. On the other hand,  for ML-WSe$_2$ PL peaks that are not in resonant condition, a reduction in the WSe$_2$ PL intensity is observed after the field induced FM order of CrSBr. This effect is associated with a change in the degree of charge transfer after the magnetic field induced phase transition of CrSBr.Furthermore, we show that this PL enhancement is anisotropic and has a short range interaction (< 2 nm). However, further studies are necessary to understand in detail the nature of different mechanisms in the WSe$_2$/CrSBr heterostructure. Our findings underscore that the magnetic control of the  RET could be an useful tool to modify the optical properties of 2D materials in anisotropic magnetic vdWHs. The appropriate design of magnetic vdWHs could be an interesting platform for the engineering of new devices for possible applications in optoelectronics, opto-spintronics and quantum technology.

\begin{suppinfo}

Details of sample preparation and complementary low temperature PL and magneto-PL results for different laser positions are shown in the Supplementary Information (PDF).

\end{suppinfo}
\begin{acknowledgement}

This work was supported by the "Fundação de Amparo a Pesquisa do Estado de São Paulo" (FAPESP, Grants Nos. 2022/10340-2, 2022/08329-0, 2023/11265-7 and 2023/01313-4) and “Conselho Nacional de Desenvolvimento Científico e Tecnológico”(CNPq, Grants Nos. 306971/2023-2, 306170/2023-0,423423/2021-5, 312705/2022-0, 2019/14017-9, 309920/2021-3, 301145/2025-3 and 151130/2023-0). YGG and SR acknowledge support from the FAPESP-SPRINT project (Grant No. 2023/08276-7). YGG and PEFJ acknowledge the financial support from “Coordenação de Aperfeiçoamento de Pessoal de Nível Superior” (CAPES)-Probal program (Grant No.88881.895140/2023-01). The authors acknowledge the Brazilian Synchrotron Light Laboratory (LNLS) for access to the Microscopic Samples Laboratory (LAM) (Proposal LAM-2D: 20240165) and the
Microscopy Atomic Force Facility (Proposal MFA: 20242480) at the Brazilian Nanotechnology National Laboratory (LNNano), which operates within the Brazilian Center for Research in Energy and Materials (CNPEM), a private nonprofit organization supervised by the Ministry of Science, Technology, and Innovations (MCTI) of Brazil. This project received funding from the European Union Horizon 2020 research and innovation program under grant agreement no. 863098 (SPRING) and Marie Sklodowska–Curie individual fellowship No. 101027187-PCSV. TSG acknowledges support from the Dutch Research Council (NWO) for a Rubicon grant (Project No. 019.222EN.013). F.D. gratefully acknowledges funding from the German Research Foundation via the Emmy Noether Program (Project-ID 534078167).

\end{acknowledgement}

\bibliography{references}

\end{document}